\documentclass{article}
\usepackage[utf8]{inputenc}
\usepackage[margin=1in, includehead, includefoot]{geometry}
\usepackage{amsthm,amssymb,amsmath}
\usepackage{enumitem}
\usepackage{mathrsfs}
\usepackage{mathtools}
\usepackage{fancyhdr}
\usepackage{indentfirst}
\usepackage{graphicx}
\usepackage{placeins}
\usepackage{wrapfig}
\usepackage{url}
\usepackage{array}

\usepackage{xcolor}
\definecolor{lightgray}{rgb}{0.83, 0.83, 0.83}
\definecolor{upforestgreen}{rgb}{0.0, 0.27, 0.13}
\definecolor{tealblue}{rgb}{0.21, 0.46, 0.53}
\definecolor{auburn}{rgb}{0.43, 0.21, 0.1}
\definecolor{blue(pigment)}{rgb}{0.2, 0.2, 0.6}
\definecolor{chocolate(traditional)}{rgb}{0.48, 0.25, 0.0}

\usepackage{lineno}

\newtheorem{definition}{Definition}
\newtheorem{lemma}{Lemma}[section]
\newtheorem{theorem}{Theorem}[section]
\newtheorem{corollary}{Corollary}[section]

\DeclarePairedDelimiter\ceil{\lceil}{\rceil}
\DeclarePairedDelimiter\floor{\lfloor}{\rfloor}

\newcommand{\toroman}[1]{\textit{\expandafter{\romannumeral #1\relax}}}

\newcommand{\cbeginproof}[0]{\par\noindent\textit{Proof.} }
\newcommand{\cendproof}[0]{ \qed\par\vspace{1em}}

\newcommand{\npcompleteproblem}[3]{ \par\vspace{0.5em}\noindent{\textbf{#1}}\newline \textbf{INSTANCE: } #2 \newline \textbf{QUESTION: } #3 \par\vspace{0.5em} }

\title{Error-correcting Identifying Codes}
\author{
    \small Devin C. Jean\\
    \small Computer Science Department \\
    \small Vanderbilt University\\
    \small \texttt{devin.c.jean@vanderbilt.edu}
    \and
    \small Suk J. Seo\\
    \small Computer Science Department\\
    \small Middle Tennessee State University\\
    \small \texttt{Suk.Seo@mtsu.edu}
}
\date{}

\begin{document}
\maketitle
\thispagestyle{empty}

\begin{abstract}
Assume that a graph $G$ models a detection system for a facility with a possible ``intruder," or a multiprocessor network with a possible malfunctioning processor.
We consider the problem of placing (the minimum number of) detectors at a subset of vertices in $G$ to automatically determine if there is an intruder, and if so, its precise location.
In this research we explore a fault-tolerant variant of identifying codes, known as error-correcting identifying codes, which permit one false positive or negative and are applicable to real-world systems.
We present the proof of NP-completeness of the problem of determining said minimum size in arbitrary graphs, and determine bounds on the parameter in cubic graphs.
\end{abstract}

\vspace{1em}


\section{Introduction}
Let $G$ be an (undirected) graph with vertices $V(G)$ and edges $E(G)$.
The \textit{open neighborhood} of a vertex $v \in V(G)$, denoted $N(v)$, is
the set of vertices adjacent to $v$, $N(v) = \{w\in V(G): vw\in E(G)\}$.
The \textit{closed neighborhood} of a vertex $v \in V(G)$, denoted $N[v]$, is $N(v) \cup \{v\}$.
If $S \subseteq V(G)$ and every vertex in $V(G)$ is within distance 1 of some $v \in S$ (i.e., $\cup_{v \in S}{N[v]} = V(G)$), then $S$ is said to be a \emph{dominating set}; for $u \in V(G)$, we let $N_S[u] = N[u] \cap S$ and $N_S(u) = N(u) \cap S$ denote the dominators of $u$ in the closed and open neighborhoods, respectively.

A set $S \subseteq V(G)$ is called a \emph{detection system} if each vertex in $S$ is installed with a specific type of detector or sensor for locating an ``intruder" such that the set of sensor data from all detectors in $S$ can be used to precisely locate an intruder, if one is present, anywhere in the graph.
Given a detection system $S \subseteq V(G)$, two distinct vertices $u,v \in V(G)$ are said to be \emph{distinguished} if it is always possible to eliminate $u$ or $v$ as the location of an intruder (if one is present).
In order to locate an intruder anywhere in the graph, every pair of vertices must be distinguished.

Many types of detection systems with various properties have been explored throughout the years, each with their own domination and distinguishing requirements.
For example, an \emph{Identifying Code (IC)} \cite{NP-complete-ic, karpovsky} is a detection system where each detector at a vertex $v \in V(G)$ can sense an intruder within $N[v]$, but does not know the exact location.
In an IC, $S$, $u$ and $v$ are distinguished if $|N_S[u] \triangle N_S[v]| \ge 1$, where $\triangle$ denotes the symmetric difference.
A \emph{Locating-Dominating (LD) set} is a detection system that extends the capabilities of an IC by allowing detectors to differentiate an intruder in $N(v)$ versus $\{v\}$ \cite{dom-loc-acyclic, ftld}.
In an LD set, $S$, $x \in S$ is automatically distinguished from all other vertices, and $u,v \notin S$ are distinguished if $|N_S[u] \triangle N_S[v]| \ge 1$.
Still another system is called an \emph{Open-Locating-Dominating (OLD) set}, where each detector at a vertex $v \in V(G)$ can sense an intruder within $N(v)$, but not at $v$ itself \cite{old, oldtree}.
In an OLD set, $S$, $u$ and $v$ are distinguished if $|N_S(u) \triangle N_S(v)| \ge 1$.
Lobstein \cite{dombib} maintains a bibliography of currently over 470 articles published on various types of detector-based sets, and other related concepts including fault-tolerant variants of ICs, LD and OLD sets.

The aforementioned detection systems assume that all detectors work properly and there are no transmission errors; for applications in real-world systems, we often desire some level of fault-tolerance built into the system.
Three common fault-tolerant properties of detection systems are \emph{Redundant Detection Systems} \cite{redic, redld, ftsets}, which allow one detector to be removed, \emph{Error-Detecting Detection Systems} \cite{ourtri, detld, ftld}, which can tolerate one false negative from a sensor, and \emph{Error-Correcting Detection Systems} \cite{our3-4, errld, ft-old-cubic}, which handle any single sensor error (a false positive or false negative).

In this paper, we will focus on Error-correcting Identifying Codes (ERR:ICs), including a full characterization and existence criteria in Section~\ref{sec:erric-char}.
For the ERR:IC parameter, ERR:IC($G$) denotes the minimum cardinality of an error-correcting IC on graph $G$.
For many detection systems and their fault tolerant variants, minimizing a detection system is known to be NP-complete for arbitrary graphs \cite{NP-complete-ic, NP-complete-ld, errld, redld, detld, redic, old}.
In Section~\ref{sec:npc}, we will prove the problem of determining ERR:IC(G) for an arbitrary graph $G$ is also NP-complete.
In Section \ref{sec:erric-cubic} we determine the bounds on value of ERR:IC($G$) for cubic graphs.

\section{Characterization and Existence Criteria of ERR:IC}\label{sec:erric-char}

Detection systems commonly use general terminology such as ``dominated" or ``distinguished", whose specific definitions vary depending on the sensors' capabilities and the level of fault-tolerance.
The following definitions are specifically for identifying codes and their fault-tolerant variants; assume that $S \subseteq V(G)$ is the set of detectors.

\begin{definition}\label{def:k-dom}
A vertex $v \in V(G)$ is \emph{$k$-dominated} by a dominating set $S$ if $|N_S[v]| = k$.
\end{definition}

\begin{definition}\label{def:k-disty}
If $S$ is a dominating set and $u,v \in V(G)$, $u$ and $v$ are \emph{$k$-distinguished} if $|N_S[u] \triangle N_S[v]| \ge k$, where $\triangle$ denotes the symmetric difference.
\end{definition}

We will also use terms such as ``at least $k$-dominated" to denote $j$-dominated for some $j \ge k$.

\vspace{0.6em}
Jean and Seo \cite{detic, redic} have shown the necessary and sufficient properties of two fault-tolerant identifying codes: redundant identifying codes (RED:ICs) and error-detecting identifying codes (DET:ICs).
Seo and Slater \cite{separating} characterized error-separating sets, which are a more general, set-theoretic form of error-correcting detection systems; we can convert their characterization to the following for error-correcting identifying codes (ERR:ICs).

\begin{theorem}[\cite{separating}]\label{theo:erric-char}
A detector set, $S \subseteq V(G)$, is an ERR:IC if and only if each vertex is at least 3-dominated and all pairs are 3-distinguished.
\end{theorem}

Table~\ref{tab:ft-ic-cmp} gives a summary of requirements for IC, redundant identifying codes (RED:ICs), error-detecting identifying codes (DET:ICs), and error-correcting identifying codes (ERR:ICs).
Before proving the existence criteria for an ERR:IC, we consider those for IC and RED:C.

\begin{table}[ht]
    \centering
    { 
        \setlength\extrarowheight{0.2em}
        \begin{tabular}{|c|c|c|}
            \hline \textbf{Detection System} & \textbf{Domination Requirement} & \textbf{Distinguishing Requirement} \\[0.2em]\hline
            IC \cite{redic} & $|N_S[u]| \ge 1$ & $|N_S[u] \triangle N_S[v]| \ge 1$ \\[0.2em]\hline
            RED:IC \cite{redic} & $|N_S[u]| \ge 2$ & $|N_S[u] \triangle N_S[v]| \ge 2$ \\[0.2em]\hline
            DET:IC \cite{detic} & $|N_S[u]| \ge 2$ & $|N_S[u] - N_S[v]| \ge 2$ or $|N_S[v] - N_S[u]| \ge 2|$ \\[0.2em]\hline
            ERR:IC (Theorem~\ref{theo:erric-char}) & $|N_S[u]| \ge 3$ & $|N_S[u] \triangle N_S[v]| \ge 3$ \\[0.2em]\hline
        \end{tabular}
    }
    \caption{Characterizations of various fault-tolerant identifying codes.}
    \label{tab:ft-ic-cmp}
\end{table}

\begin{definition}\cite{ld-twin-free} 
Two distinct vertices $u,v \in V(G)$ are said to be \emph{twins} if $N[u] = N[v]$ (\emph{closed twins}) or $N(u) = N(v)$ (\emph{open twins}).
\end{definition}

It is easy to see $G$ has an IC if and only if $G$ has no closed-twins.
Jean and Seo have proved the existence criteria for RED:IC as follows.

\begin{theorem}[\cite{redic}]\label{theo:redic-exist-3}
Let $G$ be connected with $n \ge 4$.
RED:IC exists if and only if there are no closed twins, every support vertex is at least degree three, and every triangle $abc \in G$ has $|N[a] \triangle N[b]| \ge 2$.
\end{theorem}

\begin{theorem}\label{theo:erric-exist-alt}
A graph $G$ has an ERR:IC if and only if it satisfies the following properties.
\begin{enumerate}[label=\roman*,noitemsep]
    \item $G$ is twin-free
    \item $G$ has $\delta(G) \ge 2$
    \item $G$ has no adjacent degree 2 vertices
    \item Every triangle $abc \in G$ has $|N[a] \triangle N[b]| \ge 3$
\end{enumerate}
\end{theorem}
\begin{proof}
First, we will show that these conditions are necessary for ERR:IC to exist.
Suppose for a contradiction that $G$ has an ERR:IC but fails to satisfy some of the above properties.
If $G$ has twin vertices $u,v \in V(G)$, then $u$ and $v$ cannot be distinguished, contradicting that ERR:IC exists.
If $\delta(G) < 2$ then $\exists u \in V(G)$ with $deg(u) \le 1$; thus, $u$ is not 3-dominated, a contradiction.
If there are two adjacent vertices $u,v \in V(G)$ with $deg(u) = deg(v) = 2$, then they will not be distinguished, a contradiction.
Lastly, property~\toroman{4} is directly based on the distinguishing requirements of Theorem~\ref{theo:erric-char}, and so is necessary for ERR:IC to exist.

Next, we will assume $G$ has all of the above properties and show that $S = V(G)$ is an ERR:IC for $G$.
By property~\toroman{2}, we know $\delta(G) \ge 2$, so all vertices are at least 3-dominated; we now need only show that two arbitrary vertices $u,v \in V(G)$ are distinguished.
\textbf{Case~1:} $uv \notin E(G)$.
By property~\toroman{1}, $G$ is twin-free, so without loss of generality let $x \in N(u) - N[v]$.
Then $u$ and $v$ are 3-distinguished by $u$, $v$, and $x$.
\textbf{Case~2:} $uv \in E(G)$ and $uvp$ is a triangle.
Property~\toroman{4} directly gives us that $u$ and $v$ are distinguished.
\textbf{Case~3:} $uv \in E(G)$ and $uv$ is not part of any triangle.
By property~\toroman{2}, $\delta(G) \ge 2$, so let $\exists x \in N(u) - N[v]$ and $\exists y \in N(v) - N[u]$.
By property~\toroman{3}, we know that $u$ and $v$ cannot both be degree 2, so without loss of generality assume $\exists z \in N(u) - N[v]$ with $z \neq x$.
Then $u$ and $v$ are 3-distinguished by $x$, $y$, and $z$.
Therefore, we see that $S = V(G)$ satisfies Theorem~\ref{theo:erric-char}, so $S$ is an ERR:IC for $G$.
\end{proof}

From Theorem~\ref{theo:erric-exist-alt}, we see that cycles do not have ERR:IC because they contain adjacent degree 2 vertices, and trees do not have ERR:IC because $\delta(G) \le 1$.

\begin{corollary}\label{cor:erric-exit-tri-free}
A triangle-free graph, $G$, has an ERR:IC if and only if $G$ is twin-free, $\delta(G) \ge 2$, and $G$ has no two adjacent degree 2 vertices.
\end{corollary}

\begin{figure}[ht]
    \centering
    \includegraphics[width=0.4\textwidth]{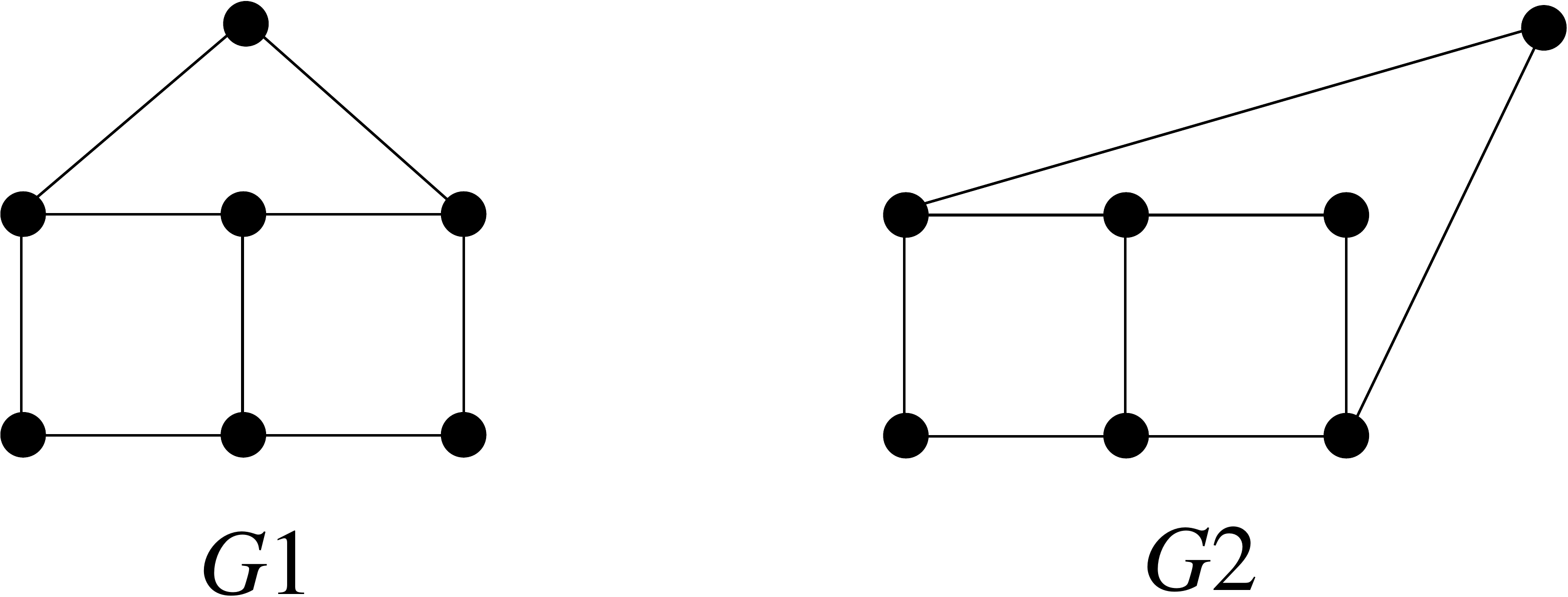}
    \caption{The two graphs supporting ERR:IC which have smallest $n$}
    \label{fig:g7}
\end{figure}

\begin{theorem}\label{theo:erric-g7-1-2-unique}
If $G$ is a graph with ERR:IC and $n \le 7$, then $G = G1$ or $G = G2$, as shown in Figure~\ref{fig:g7}.
\end{theorem}
\begin{proof}
We know that $G$ cannot be acyclic because trees have $\delta(G) \le 1$, contradicting Theorem~\ref{theo:erric-exist-alt} property~\toroman{2}.
We will proceed by casing on the existence of cycles of a given size.

Case 1: $G$ has a 3-cycle $abc$.
Suppose $deg(a) = 2$.
Let $B = N[b] - N[a]$ and $C = N[c] - N[a]$.
From Theorem~\ref{theo:erric-char}, distinguishing $(a,b)$ and $(a,c)$ requires $|B| \ge 3$ and $|C| \ge 3$, and $n \le 7$ forces $|B \cup C| \le 4$.
This implies that $|B \cap C| \ge 2$, meaning $|N[b] \triangle N[c]| = |B \triangle C| = |B \cup C| - |B \cap C| \le 2$, contradicting that $G$ has an ERR:IC.
Otherwise, by symmetry, we assume all vertices of $abc$ are at least degree 3.
Suppose $deg(a) = 6$; then we already have $n = 7$ and cannot add more vertices.
To distinguish $b$ and $c$, without loss of generality we can assume that $|N[b] - \{a,b,c\}| \ge 2$.
We see that $a$ and $b$ cannot be distinguished, a contradiction.
Otherwise we assume all vertices in $abc$ have degrees between 3 and 5.
Suppose $deg(a) = 5$; let $\{x,y,z\} = N(a) - \{a,b,c\}$.
By symmetry, we can assume $|N(b) \cap \{x,y,z\}| \ge |N(c) \cap \{x,y,z\}|$.
To distinguish $b$ and $c$, we need $|N(b) \cap \{x,y,z\}| \ge 1$.
To distinguish $a$ and $b$, we require $n \ge 7$ with a new vertex $w \in N(b)$.
If $N(c) \cap \{x,y,z\} \neq \varnothing$ then by similar logic we require $w \in N(c)$ to distinguish $a$ and $c$, but this results in $b$ and $c$ being impossible to distinguish; therefore, we assume $N(c) \cap \{x,y,z\} = \varnothing$.
If $w \in N(c)$, then $b$ and $c$ cannot be distinguished, a contradiction.
Otherwise $w \notin N(c)$, meaning $deg(c) = 2$, a contradiction.
Now by symmetry we can assume each vertex in $abc$ has degree 3 or 4.
Suppose $deg(a) = 4$, and let $\{x,y\} = N(a) - \{a,b,c\}$.
By symmetry, we assume that $|N(b) \cap \{x,y\}| \ge |N(c) \cap \{x,y\}|$.
To distinguish $b$ and $c$, we require $|N(b) \cap \{x,y\}| \ge 1$.
If $|N(b) \cap \{x,y\}| = 2$, then $a$ and $b$ are not distinguished, so we assume $|N(b) \cap \{x,y\}| = 1$.
To distinguish $a$ and $b$, we require $n = 7$ with two new vertices, $p$ and $q$, with $p,q \in N(b)$.
If $N(c) \cap \{x,y\} \neq \varnothing$ then distinguishing $a$ and $c$ would require $p,q \in N(c)$, but then $b$ and $c$ cannot be distinguished; therefore, we can assume $N(c) \cap \{x,y\} = \varnothing$.
If $N(c) \cap \{p,q\} \neq \varnothing$, then $b$ and $c$ cannot be distinguished, so we assume $N(c) \cap \{p,q\} = \varnothing$.
Thus, $deg(c) = 2$, a contradiction.
Now, by symmetry we can assume all vertices in $abc$ are degree 3.
We observe that no two vertices in $abc$ can be distinguished, a contradiction.

Case 2: $G$ has a 4-cycle $abcd$; from the previous case, we can assume $G$ is triangle-free.
If $abcd$ has two adjacent degree-2 vertices, then it would violate Theorem~\ref{theo:erric-exist-alt} property~\toroman{3}, a contradiction.
If $abcd$ has two opposite vertices of degree 2, then they would be twins, contradicting Theorem~\ref{theo:erric-exist-alt} property~\toroman{1}.
Therefore, without loss of generality we can assume $a$, $b$, $c$ have at least degree 3, and call these vertices $a',b',c'$ with $a' \in N(a)$, $b' \in N(b)$, and $c' \in N(c)$.
We know that $a' \neq b'$ and $b' \neq c'$ because $G$ is triangle-free.
Suppose $a' = c'$.
If $deg(a) = 3 = deg(c)$, then $a$ and $c$ are twins, a contradiction; otherwise, without loss of generality, let $deg(a) \ge 4$ and call the new vertex $w \in N(a)$.
If $deg(d) = 2 = deg(a')$, then $d$ and $a'$ are twins, a contradiction, so without loss of generality let $deg(a') \ge 3$; there are already $n=7$ vertices, so let $a'b' \in E(G)$, which is the only edge that can be added to $a'$ without creating a triangle.
We find that $a'$ and $b$ are twins, a contradiction.
Now, we can assume $a' \neq c'$.
If $deg(d) = 3$, then $db' \in E(G)$ is required, and $b$ and $d$ would be twins, a contradiction; thus, we can assume $deg(d) = 2$.
We know that $\delta(G) \ge 2$, so without loss of generality let $a'b' \in E(G)$ to make $deg(b') \ge 2$.
If $b'c' \in E(G)$ then we arrive at $G1$, and we note that no more edges can be added without violating the existence of ERR:IC; otherwise, we assume $b'c' \notin E(G)$.
To make $deg(c') \ge 2$, we require $ac' \in E(G)$ or $a'c' \in E(G)$.
If $a'c' \in E(G)$, then we arrive at $G2$, and we note that no more edges can be added without violating the existence of ERR:IC.
Otherwise $a'c' \notin E(G)$, so $ac' \in E(G)$ is required.
We see that $d$ and $c'$ are twins, a contradiction.

Case 3: $G$ has a $k$-cycle, $C$, for $5 \le k \le 7$.
From previous cases, we can assume $G$ has girth $k$, implying no chords can be added in $C$.
Due to Theorem~\ref{theo:erric-exist-alt} property~\toroman{3}, there cannot be adjacent degree 2 vertices---implying degree 2 vertices must form an independent set on $C$---so we require $n' \ge \ceil{\frac{k}{2}} \ge 3$ vertices in $C$ to have at least degree 3.
However, because $n \le 7$, we can only add $n'' = 7 - k \le 2$ new vertices.
If $n'' = 0$, then $G = C_7$ and there are adjacent degree 2 vertices, a contradiction.
Otherwise, by the pigeon hole principle, there must be two distinct vertices $p,q \in V(C)$ with a common $w \in (N(p) \cap N(q)) - V(C)$, which will form a cycle with length $\ell = d+2$ where $d = d_C(p,q)$ is the distance between $p$ and $q$ along the cycle $C$.
We know that $\ell = d + 2 \le \floor{\frac{k}{2}} + 2 = \floor{k - \frac{k}{2}} + 2 = k - \ceil{\frac{k}{2}} + 2$.
Because $k \ge 5$, $\ell < k$, a contradiction.
\end{proof}

Because non-detectors provide no utility, we know that if $S \subseteq V(G)$ is an ERR:IC for $G$, then $G[S]$ (the graph induced by $S$) has an ERR:IC, namely $S$.
Therefore, $G1$ and $G2$ from Figure~\ref{fig:g7} have $\textrm{ERR:IC}(G)=7$.

\begin{corollary}
If $G$ has an ERR:IC, $S$, then $n \ge |S| \ge 7$.
\end{corollary}

From Theorem~\ref{theo:erric-exist-alt}, we see that if a cubic (3-regular) graph has a triangle it will violate property~\toroman{4}.
Thus, we have the following corollary.

\begin{corollary}\label{cor:erric-exist-cubic}
A cubic graph, $G$, has an ERR:IC if and only if it is twin-free and triangle-free.
\end{corollary}

We observe that a 0- or 1-regular graph will have $\delta(G) < 2$, contradicting Theorem~\ref{theo:erric-exist-alt} property~\toroman{2}, and a 2-regular graph will violate property~\toroman{3}.
In general, we have the following corollary.

\begin{corollary}
A $k$-regular graph, $G$, has an ERR:IC if and only if $k \ge 3$, $G$ is twin-free, and any triangle $abc \in G$ has $|N[a] \triangle N[b]| \ge 3$.
\end{corollary}

We conclude this section with the following two theorems that relate existence criteria between several fault-tolerant variants of ICs.

\begin{theorem}\label{theo:kreg-ic-iff-redic}
If $G$ is $k$-regular for $k \ge 2$, then IC exists if and only if RED:IC exists.
\end{theorem}
\begin{proof}
We know that existence of RED:IC implies existence of IC, so we need only show the converse.
Let $G$ be a $k$-regular graph for which IC exists, and let $u,v \in V(G)$ be distinct vertices; we will show $S = V(G)$ is a RED:IC.
Because $deg(u) = deg(v) = k$, we know $|(N[u] \cap S) \triangle (N[v] \cap S)| = |N[u] \triangle N[v]| = (k+1) + (k+1) - 2|N[u] \cap N[v]| = 2j$ for some $j \in \mathbb{N}_0$.
We know $j \neq 0$ because IC is assumed to exist, so $j \ge 1$, meaning all vertices are at least 2-distinguished.
Because $G$ is $k$-regular for $k \ge 2$, every vertex is at least 2-dominated; thus, $S$ is a RED:IC, completing the proof.
\end{proof}

\begin{theorem}
If $G$ is $k$-regular for $k \ge 2$, then DET:IC exists if and only if ERR:IC exists.
\end{theorem}
\begin{proof}
Similar to Theorem~\ref{theo:kreg-ic-iff-redic}, we will show that existence of DET:IC implies existence of ERR:IC.
Let $G$ be a $k$-regular graph which has DET:IC, and let $u,v \in V(G)$ be distinct vertices; we will show $S = V(G)$ is an ERR:IC.
From the proof of Theorem~\ref{theo:kreg-ic-iff-redic}, we know $|(N[v] \cap S) \triangle (N[u] \cap S)| = 2j$ for some $j \in \mathbb{N}_0$.
We know $j \neq 0$ because DET:IC exists.
We also know $j \neq 1$ because $deg(u) = deg(v)$ and $j=1$ would imply that $u$ and $v$ are only 2-distinguished rather than the $2^\#$-distinguishing required by DET:IC.
Thus, $j \ge 2$, meaning all vertices are at least 3-distinguished, and all vertices are at least 3-dominated because $k \ge 2$.
Therefore, $S$ is an ERR:IC, completing the proof.
\end{proof}

\section{NP-completeness of ERR:IC}\label{sec:npc}

Many graphical parameters related to detection systems, such as finding optimal IC, LD, or OLD sets, are NP-complete problems \cite{ld-ic-np-complete-2, NP-complete-ic, NP-complete-ld, old}.
We will show that ERR-IC, the problem of determining the smallest ERR:IC set, is also NP-complete.
For additional information about NP-completeness, see Garey and Johnson \cite{np-complete-bible}.

Clearly, ERR-IC is in NP, as every possible candidate solution can be generated nondeterministically in polynomial time, and each candidate can be verified in polynomial time using Theorem~\ref{theo:erric-char}.
To show that ERR-IC is NP-complete, we will demonstrate a reduction from 3-SATISFIABILITY (3SAT) to ERR-IC.

\npcompleteproblem{3SAT}{Let $X$ be a set of $N$ variables.
Let $\psi$ be a conjunction of $M$ clauses, where each clause is a disjunction of three literals from distinct variables of $X$.}{Is there is an assignment of values to $X$ such that $\psi$ is true?}

\npcompleteproblem{Error-correcting Identifying Code (ERR-IC)}{A graph $G$ and integer $K$.}{Is there an ERR:IC set $S$ with $|S| \le K$? Or equivalently, is ERR:IC($G$) $\le K$?}

\begin{figure}[ht]
    \centering
    \includegraphics[width=0.75\textwidth]{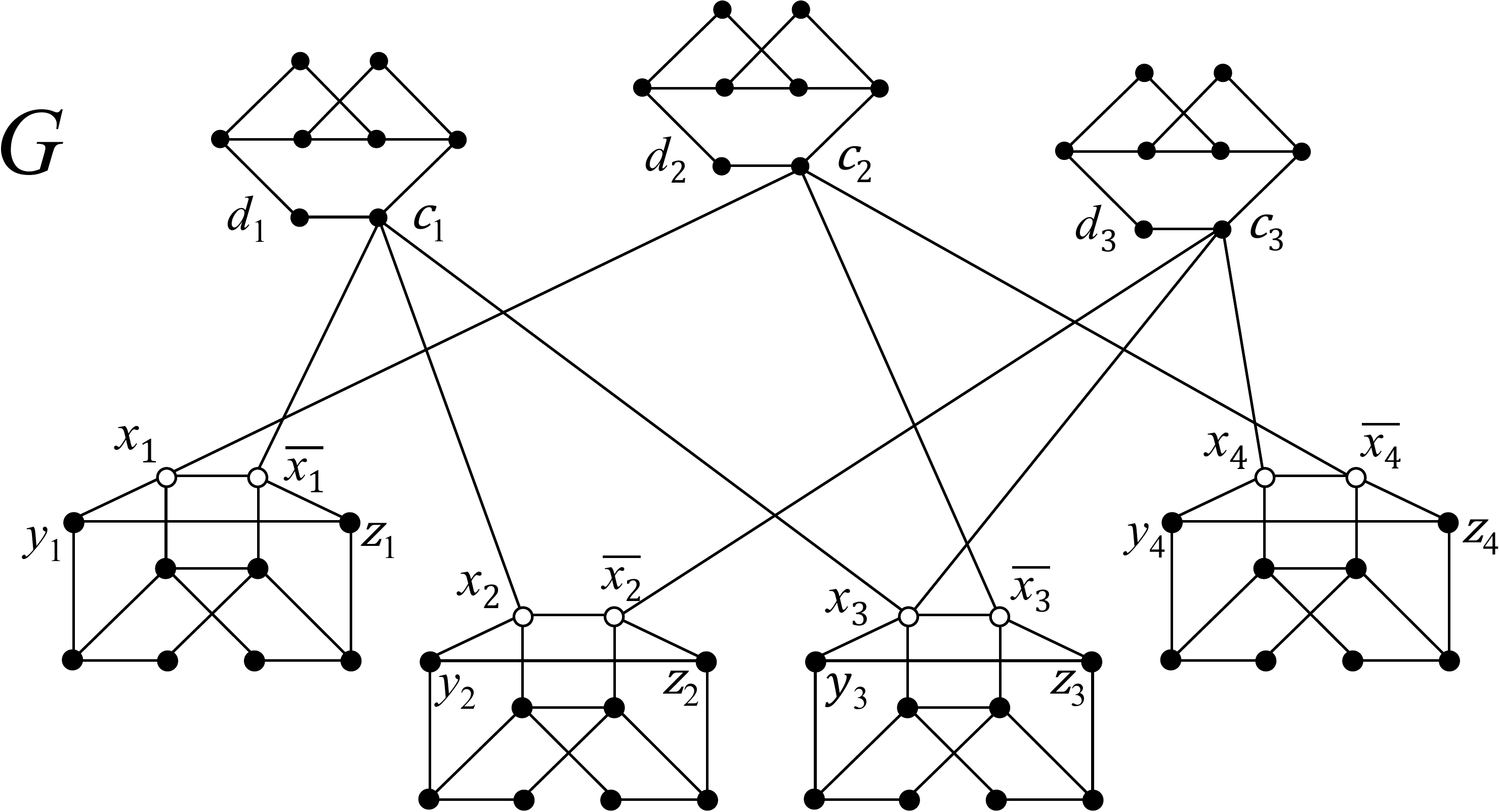}
    \caption{Example clauses: $(\overline{x_1} \lor x_2 \lor x_3) \land (x_1 \lor \overline{x_3} \lor \overline{x_4}) \land (\overline{x_2} \lor x_3 \lor x_4)$}
    \label{fig:example-clauses-err-ic}
\end{figure}

\begin{theorem}
The ERR-IC problem is NP-complete.
\end{theorem}
\cbeginproof
\begin{wrapfigure}{r}{0.37\textwidth}
    \centering
    \includegraphics[width=0.35\textwidth]{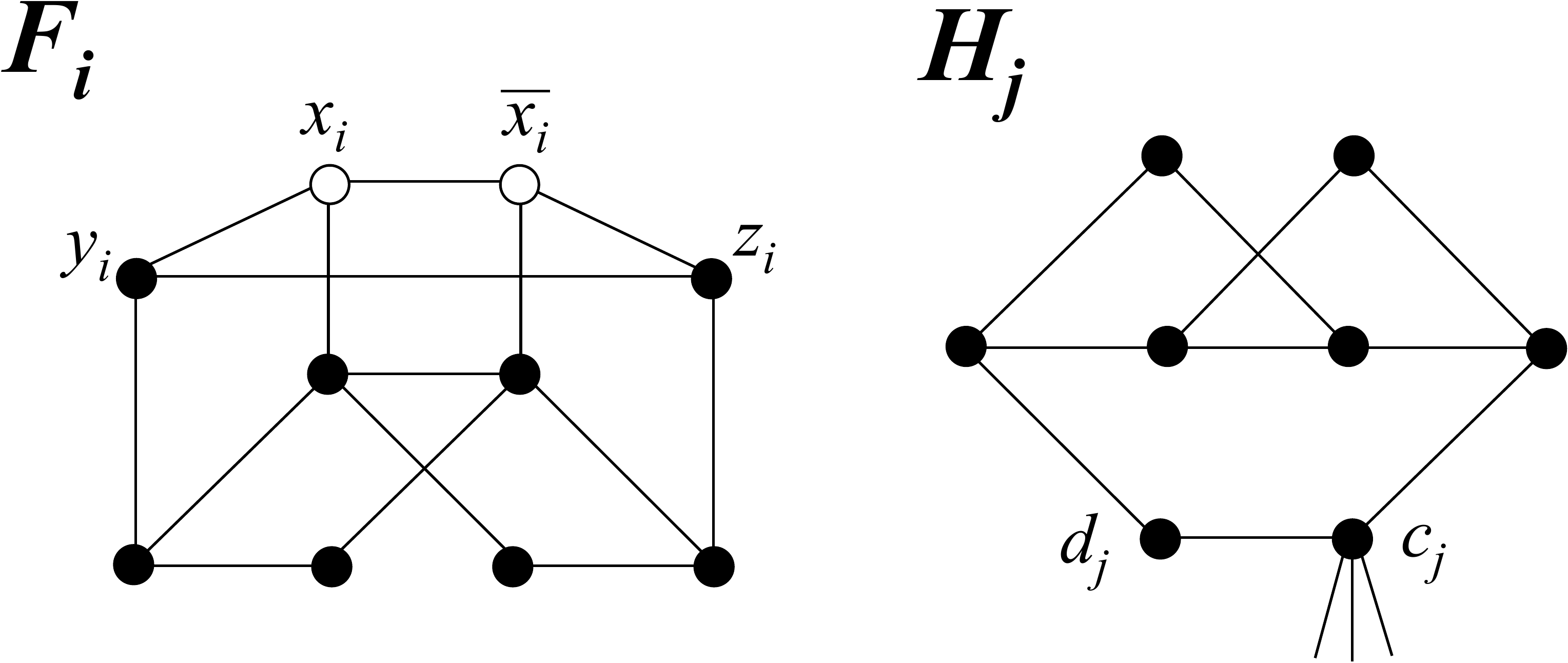}
    \caption{Variable and Clause graphs}
    \label{fig:variable-clause}
\end{wrapfigure}
Let $\psi$ be an instance of the 3SAT problem with $M$ clauses on $N$ variables.
We will construct a graph, $G$, as follows.
For each variable $x_i$, create a copy of the $F_i$ graph (Figure~\ref{fig:variable-clause}); this includes a vertex for $x_i$ and its negation $\overline{x_i}$.
For each clause $c_j$ of $\psi$, create a copy of the $H_j$ graph (Figure~\ref{fig:variable-clause}).
For each clause $c_j = \alpha \lor \beta \lor \gamma$, create an edge from the $c_j$ vertex to $\alpha$, $\beta$, and $\gamma$ in the variable graphs, each of which is either some $x_i$ or $\overline{x_i}$; for example, see Figure~\ref{fig:example-clauses-err-ic}.
The resulting graph has precisely $10N + 8M$ vertices and $15N + 13M$ edges, and can be constructed in polynomial time.

Suppose $S \subseteq V(G)$ is an optimal ERR:IC on $G$.
By Theorem~\ref{theo:erric-char}, every vertex must be 3-dominated; thus, we require at least $8N + 8M$ detectors, as shown by the shaded vertices in Figure~\ref{fig:variable-clause}.
For each $H_j$, we see that $c_j$ and $d_j$ are not distinguished unless $c_j$ is adjacent to at least one additional detector vertex.
Similarly, in each $F_i$ we see that $y_i$ and $z_i$ are not distinguished unless $\{x_i,\overline{x_i}\} \cap S \neq \varnothing$.
Thus, we find that $|S| \ge 9N + 8M$; if $|S| = 9N+8M$, then for all $i$ and $j$, $|\{x_i,\overline{x_i}\} \cap S| = 1$ and $c_j$ must be dominated by one of its three neighbors in the $F_i$ graphs, so $\psi$ is satisfiable.

Next, assume $\Psi$ is an assignment of truth values to the variables such that $\psi$ is true.
Let $S$ be the set of $8N + 8M$ detectors that are required for 3-domination.
For each variable $x_i$, if $\Psi(x_i)$ is true then we add vertex $x_i$ to $S$; otherwise, we add vertex $\overline{x_i}$ to $S$. 
Each added $x_i$ or $\overline{x_i}$ will make $y_i$ and $z_i$ distinguished and we have $|S| = 9N + 8M$.
Because $\Psi$ is a satisfying assignment for $\psi$, each $c_j$ must be adjacent to at least one additional detector vertex in the $F_i$ graphs.
Hence, $c_j$ and $d_j$ are distinguished, which makes $S$ an ERR:IC.
Therefore, $G$ has an ERR:IC of size $9N + 8M$ if and only if $\psi$ is satisfiable, completing the proof.
\cendproof

\FloatBarrier
\section{ERR:IC in Cubic Graphs}\label{sec:erric-cubic}

\begin{definition}
For $u,v \in V(G)$, the distance (length of shortest path) between $u$ and $v$ is denoted $d(u,v)$.
\end{definition}

\begin{definition}
For $v \in V(G)$, we denote $B_r(v) = \{u \in V(G) : d(u,v) \le r\}$ to be the ball of radius $r$ about $v$.
\end{definition}



\subsection{ERR:IC on the Infinite Ladder Graph}

\begin{figure}[ht]
    \centering
    \includegraphics[width=0.7\textwidth]{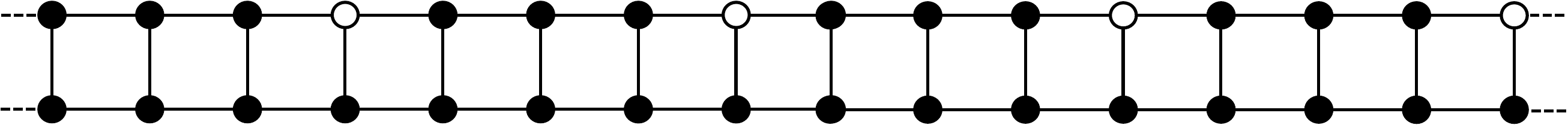}
    \caption{An optimal solution for the infinite ladder with $\textrm{ERR:IC\%}(P_\infty \square P_2) = \frac{7}{8}$}
    \label{fig:ladder-err-ic-soln}
\end{figure}

\begin{theorem}
The infinite ladder graph has $\textrm{ERR:IC\%}(P_\infty \square P_2) = \frac{7}{8}$.
\end{theorem}
\cbeginproof
The construction given by Figure~\ref{fig:ladder-err-ic-soln} is a density $\frac{7}{8}$ ERR:IC on the infinite ladder graph.
We will prove that $\frac{7}{8}$ is the optimal value by showing an arbitrary non-detector vertex can be associated with at least seven detectors.
For $v \in V(G)$, let $R_{10}(v) = B_2(v) \cup \{u \in V(G) : |N(u) \cap B_2(v)| = 2\}$.
We impose that $x$ can be associated only with detector vertices within $R_{10}(x)$.
We will allow partial ownership of detectors, so a detector vertex, $v \in S$, contributes $\frac{1}{k}$, where $k = |R_{10}(v) \cap \overline{S}|$, toward the required total of three detectors.

\begin{figure}[ht]
    \centering
    \includegraphics[width=0.4\textwidth]{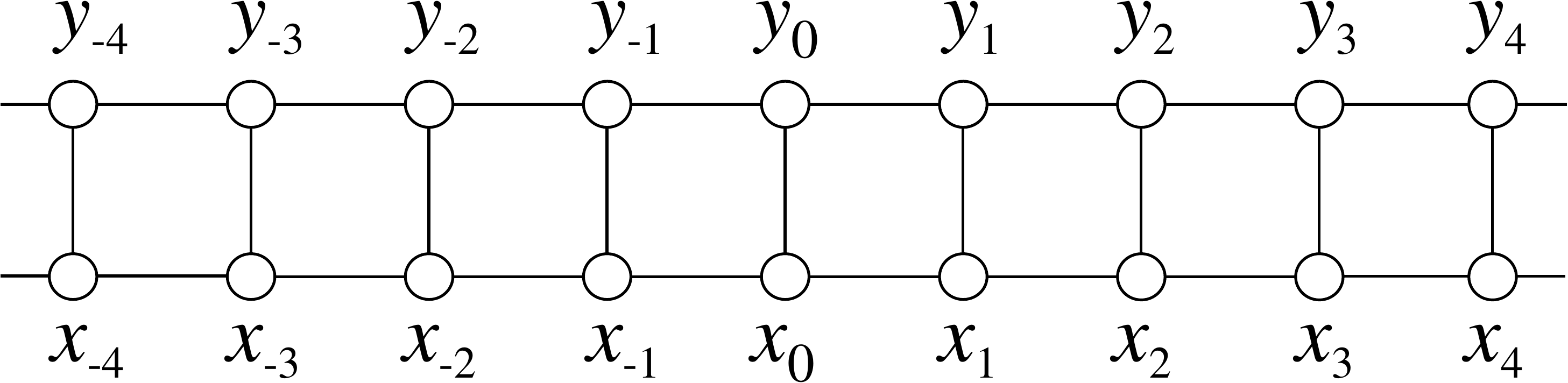}
    \caption{Ladder graph labeling scheme}
    \label{fig:ladder-labeling}
\end{figure}

Let $x_0 \notin S$ (see Figure~\ref{fig:ladder-labeling}).
To 3-dominate $x_0$, we require $\{x_{-1},x_1,y_0\} \subseteq S$.
To distinguish $x_0$ and $y_1$, we require $\{y_1,y_2\} \subseteq S$, and by symmetry $\{y_{-1},y_{-2}\} \subseteq S$.
To distinguish $x_1$ and $y_1$, we need $x_2 \in S$, and by symmetry $x_{-2} \in S$.
To distinguish $x_1$ and $x_2$, we need $x_3 \in S$ and by symmetry $x_{-3} \in S$.
Finally, to distinguish $x_1$ and $y_2$, we require $y_3 \in S$, and by symmetry  $y_{-3} \in S$.
Allowing for the possibility of $\{x_2,y_2\}$ and $\{x_{-2},y_{-2}\}$ being shared multiple times by some non-detector in $\{x_4,y_4\}$ and $\{x_{-4},y_{-4}\}$ (at most one each), we see that $x_0$ is associated wih $\frac{2}{2} + \frac{2}{2} + \frac{5}{1} = 7$ detectors, completing the proof.
\cendproof

\FloatBarrier
\subsection{ERR:IC on the Infinite Hexagonal Grid}

\begin{figure}[ht]
    \centering
    \includegraphics[width=0.45\textwidth]{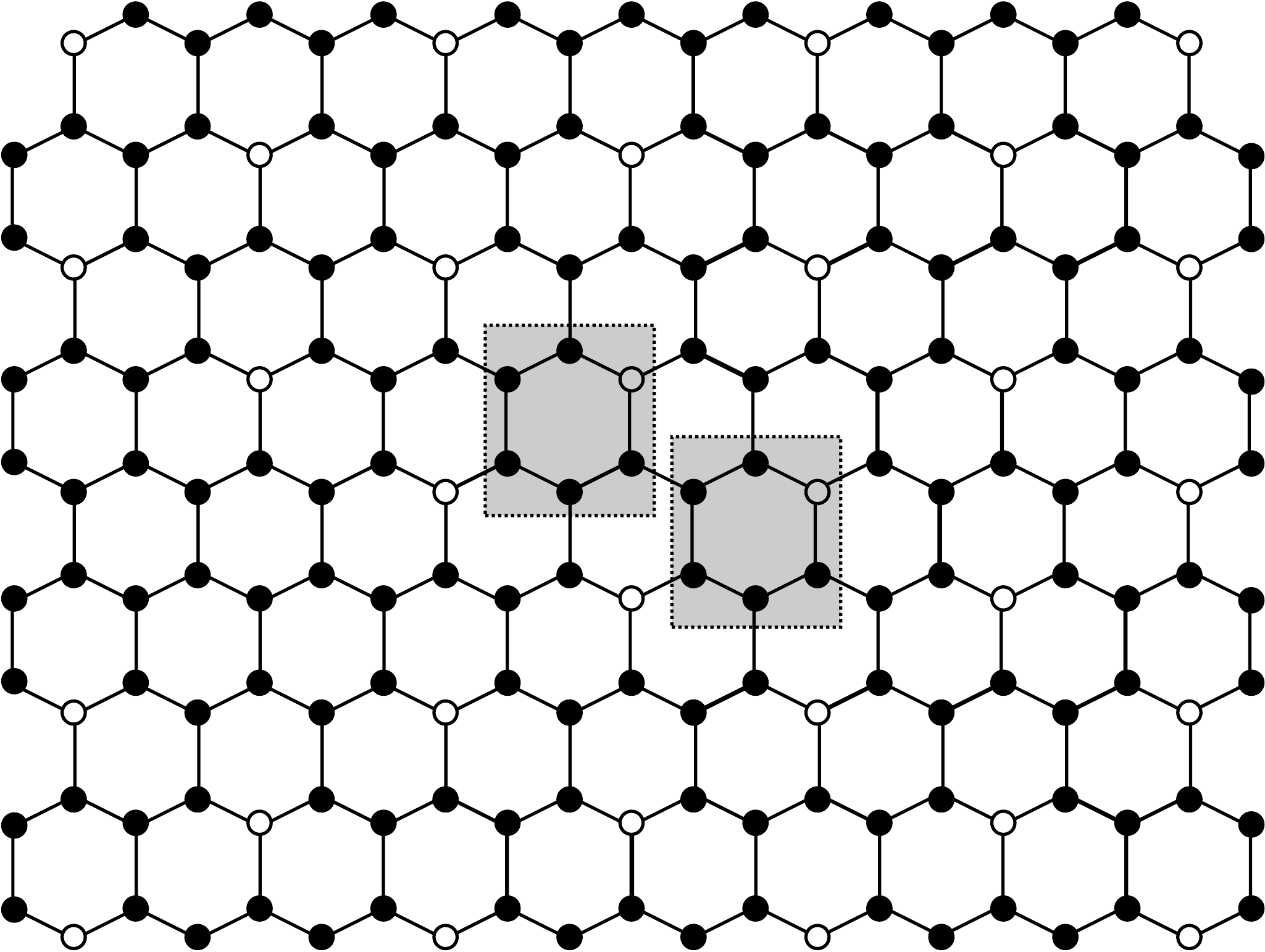}
    \caption{$\textrm{ERR:IC\%}(HEX) \leq \frac{5}{6}$}
    \label{fig:err-ic-hex}
\end{figure}

\begin{theorem}
For the infinite hexagonal grid, HEX, $\textrm{ERR:IC\%}(HEX) = \frac{5}{6}$.
\end{theorem}
\begin{proof}
Figure~\ref{fig:err-ic-hex} provides an ERR:IC for HEX which achieves density $\frac{5}{6}$, implying $\textrm{ERR:IC\%}(HEX) \le \frac{5}{6}$. This density will be proven optimal by Theorem~\ref{theo:erric-cubic-lower}, which establishes that $\textrm{ERR:IC\%}(G) \ge \frac{5}{6}$ for any cubic graph $G$.
\end{proof}

\subsection{Lower Bound on ERR:IC(G) for Cubic Graphs}

For a dominating set $S \subseteq V(G)$ of $G$ and a vertex $v \in S$, Slater \cite{ftld} defines the \emph{share} of $v$ to be $sh(v) = \sum_{u \in N[v]}{1/|N[u] \cap S|}$; that is, $v$'s contribution to the domination of its neighbors.
Each vertex $u \in V(G)$ with $|N[u] \cap S| = k$ contributes $\frac{1}{k}$ to $sh(x)$ for each $x \in N[u] \cap S$ (and $0$ to any other vertex).
Therefore, because $S$ is a dominating set, $\sum_{v \in S}{sh(v)} = n$, implying that the inverse of the average share is equal to the density of $S$ in $V(G)$.
Therefore, an upper bound on the average share (over all detectors) can be reciprocated to give a lower bound for the density.
As a shorthand, we will let $\sigma_A$  denote $\sum_{k \in A}{\frac{1}{k}}$ for some sequence of single-character symbols, $A$.
Thus, $\sigma_a = \frac{1}{a}$, $\sigma_{ab} = \frac{1}{a} + \frac{1}{b}$, and so on.
We also let $dom(v) = |N[v] \cap S|$ denote the \emph{domination number} of some vertex $v \in V(G)$.

\begin{theorem}\label{theo:erric-cubic-lower}
If $G$ is a cubic graph, then $\textrm{ERR:IC\%}(G) \ge \frac{5}{6}$.
\end{theorem}
\begin{proof}
Let $S$ be an ERR:IC for $G$.
We will show that the average share value of any detector vertex, $x \in S$ is at most $\frac{6}{5}$.
Let $N(x) = \{a,b,c\}$.
We have 2 cases to consider: $x$ is 3- or 4-dominated.

Suppose $x$ is 3-dominated; without loss of generality, let $c \notin S$.
To distinguish $x$ and $a$, we require $a$ be 4-dominated; similarly, $b$ must also be 4-dominated.
Thus, $sh(x) = \sigma_{4433} = \frac{7}{6}$.

Otherwise, $x$ is 4-dominated, meaning $\{a,b,c\} \subseteq S$.
If any of $a,b,c$ is 4-dominated, then $sh(x) \le \sigma_{4433}$ and we would be done; thus, we assume that $a,b,c$ are all 3-dominated.
From the previous case, we know that $a,b,c$ all have share $\frac{7}{6}$, but currently $sh(x)$ could be as high as $\sigma_{4333} = \frac{5}{4} > \frac{6}{5}$.
We will discharge some of $x$'s excess share into each of $a,b,c$.
Each can accept an additional $\frac{6}{5}-\frac{7}{6} = \frac{1}{30}$ total share, and each is 3-dominated, so there are potentially 2 discharge sources; thus, each can accept at most $\frac{1}{2}\frac{1}{30} = \frac{1}{60}$ additional share from each source.
Then the share of $x$ after discharging is $\frac{5}{4} - 3 \times \frac{1}{60} = \frac{6}{5}$, completing the proof.
\end{proof}

\begin{figure}[ht]
    \centering
    \includegraphics[width=0.175\textwidth]{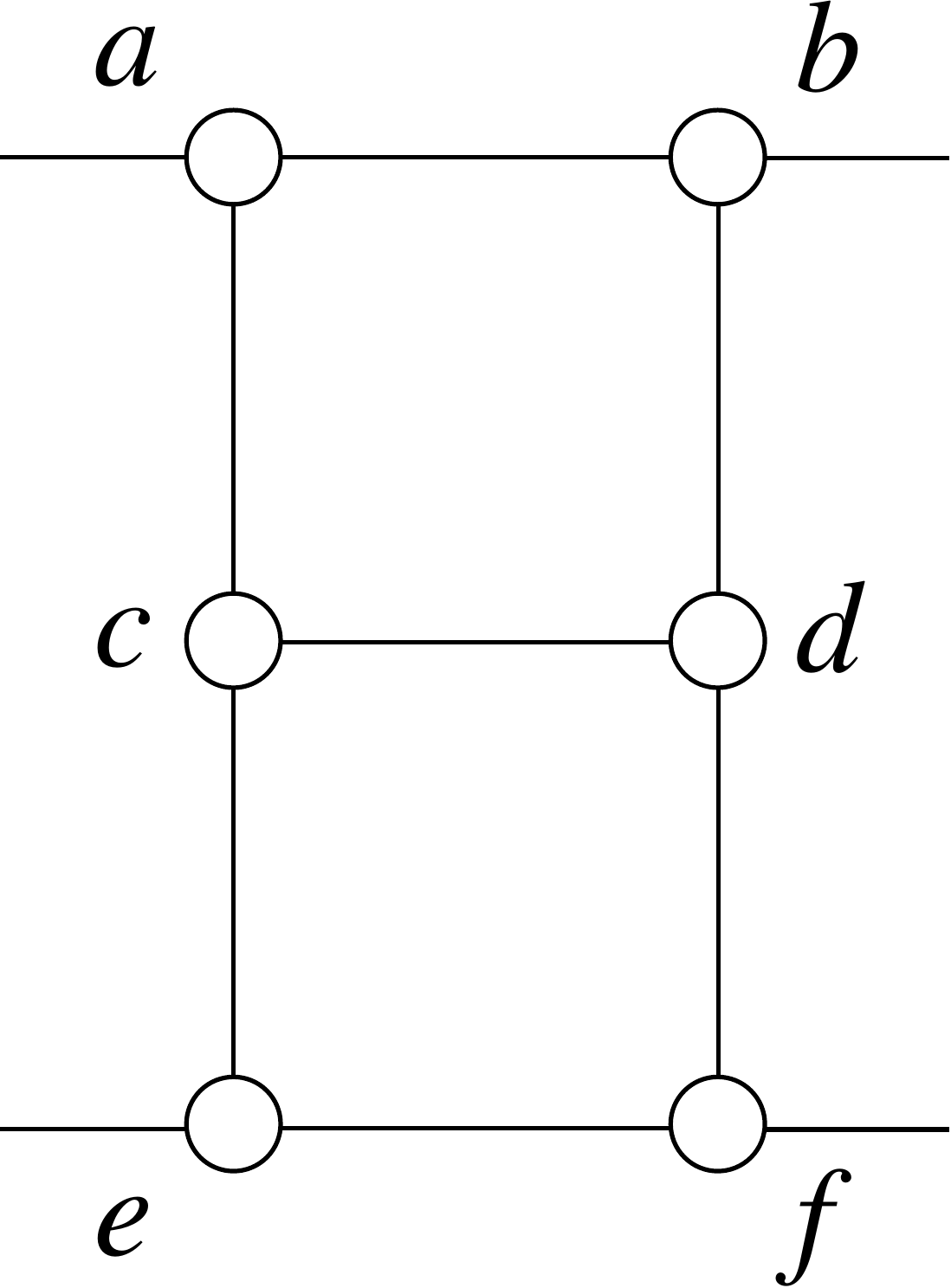}
    \caption{Subgraph $G_6$}
    \label{fig:g6}
\end{figure}

\begin{figure}[ht]
    \centering
    \includegraphics[width=0.35\textwidth]{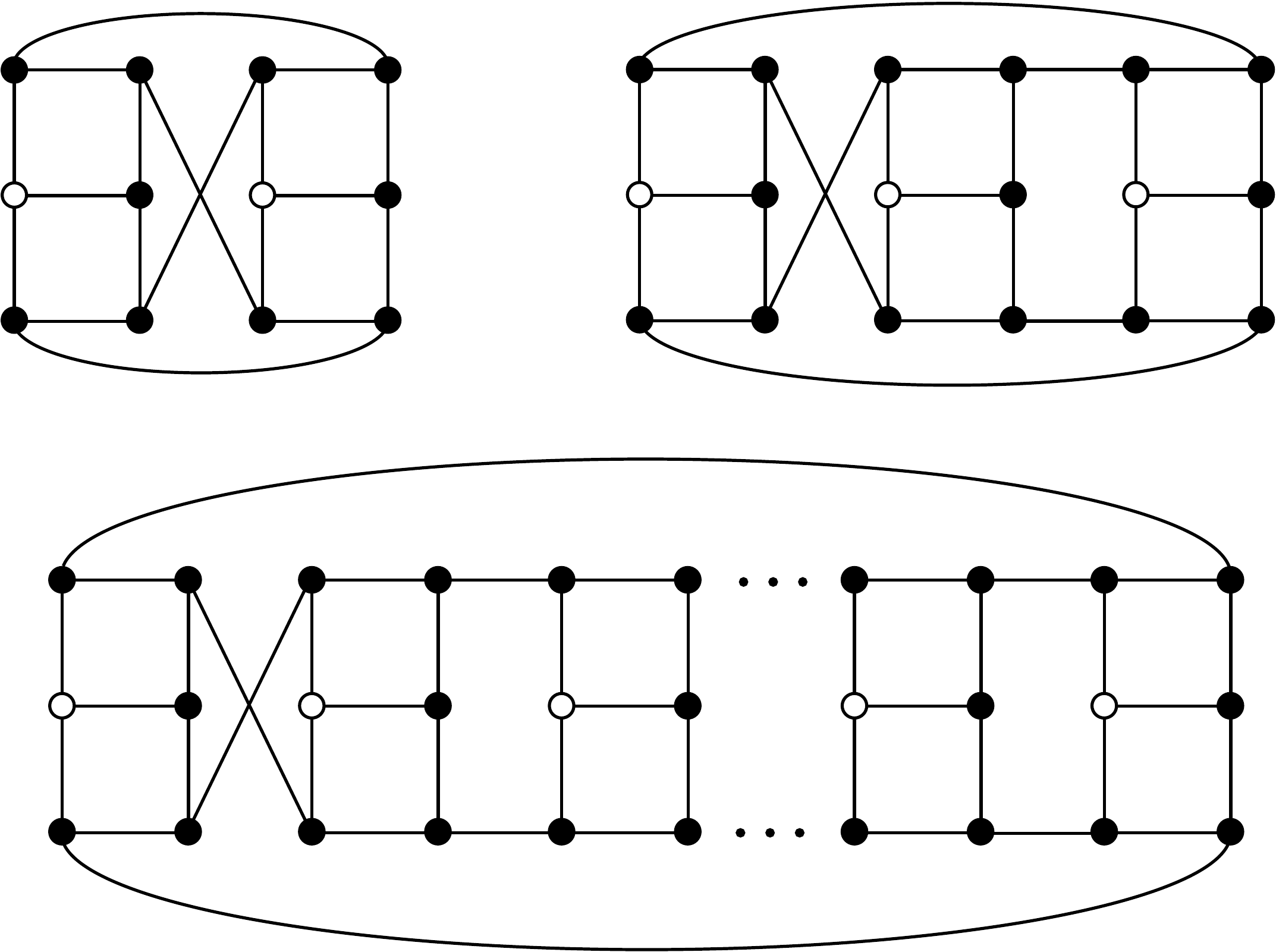}
    \caption{Infinite family of cubic graphs with ERR:IC = 5/6}
    \label{fig:cubic-fam-err-ic-lb}
\end{figure}

\FloatBarrier
\begin{theorem}
The infinite family of cubic graphs given in Figure~\ref{fig:cubic-fam-err-ic-lb} has $\textrm{ERR:IC\%}(G) = \frac{5}{6}$.
\end{theorem}
\begin{proof}
The family is constructed by connecting $k \ge 2$ copies of subgraph $G_6$ in a m\"obius ladder shape, shown in Figure~\ref{fig:g6}.
We see that $G_6$ has diameter 3, so Lemma~\ref{lem:erric-b3} yields that there can be at most one non-detector in each copy of $G_6$.
In the full graph, $G$, let $C$ denote the $k$ copies of the $c$ vertex in each copy of subgraph $G_6$.
We see that $\overline{S} = C$ satisfies Theorem~\ref{theo:erric-sbar}, so we have an ERR:IC on $G$ with density $\frac{5}{6}$.
\end{proof}

\FloatBarrier
\subsection{Upper bound on ERR:IC(G) in cubic graphs}

\begin{lemma}\label{lem:erric-b3}
If $S$ is an ERR:IC for cubic graph $G$ and $v \in V(G)-S$, then $B_3(v)-\{v\} \subseteq S$.
\end{lemma}
\begin{proof}
Let $v \in V(G)-S$.
To 3-dominate each $u \in N(v)$, we need $N[u] - \{v\} \subseteq S$; this implies that $B_2(v)-\{v\} \subseteq S$.
We now need only show that any $u \in V(G)$ with $d(u,v) = 3$ is required to be a detector.
Because $d(u,v) = 3$, we can let ($v$, $x$, $y$, $u$) be a path of length 3.
We see that $u \in S$ is required to distinguish vertices $x$ and $y$, completing the proof.
\end{proof}

\begin{lemma}\label{lem:erric-c4-1-nondet}
Let $S$ be an ERR:IC for a cubic graph $G$, let ($a$, $b$, $c$, $d$, $a$) be a 4-cycle in $G$, and let $e,f,g,h$ be adjacent to $a,b,c,d$, respectively with $\{a,b,c,d\} \cap \{e,f,g,h\} = \varnothing$. Then $|\{a,b,c,d,e,f,g,h\} \cap S| \ge 7$.
\end{lemma}
\begin{proof}
Because $S$ is an ERR:IC on $G$, it must be that $A = \{a,b,c,d,e,f,g,h\}$ are all distinct, as otherwise we produce triangles or twins.
Without loss of generality assume $\{ae,bf,cg,dh\} \in E(G)$.
If $a \notin S$, then Lemma~\ref{lem:erric-b3} yields that $A-\{a\} \subseteq B_3(a)-\{a\} \subseteq S$ and we would be done; otherwise by symmetry we assume $\{a,b,c,d\} \subseteq S$.
If $\{e,f,g,h\} \subseteq S$ we would be done, so without loss of generality assume $e \notin S$.
Lemma~\ref{lem:erric-b3} then requires that $\{f,h\} \subseteq B_3(e)-\{e\} \subseteq S$.
Finally, if $g \notin S$ then $a$ and $c$ would not be distinguished, a contradiction, completing the proof.
\end{proof}

\begin{definition}
Two vertices $p,q \in V(G)$ are called ``rivals" if there exists a 4-cycle $paqb$. And $p',q' \in (N(p) \cup N(q)) - \{p,a,q,b\}$ are called their ``friends".
\end{definition}

\begin{theorem}\label{theo:erric-sbar}
Let $G$ be a cubic graph for which ERR:IC exists. $S \subseteq V(G)$ is an ERR:IC for $G$ if and only if $\overline{S} = V(G) - S$ has $d(u,v) \ge 4$ for any distinct $u,v \in \overline{S}$ and for any rivals $p,q \in V(G)$ with friends $p',q' \in V(G)$, $|\{p',q'\} \cap \overline{S}| \le 1$.
\end{theorem}
\begin{proof}
Let $\overline{S}$ be defined as above.
Due to the distance 4 requirement between non-detectors, we know that every vertex is at least 3-dominated.
Thus, to show $S$ is an ERR:IC, we need only show that any two distinct $u,v \in V(G)$ are distinguished.

Suppose $d(u,v) = 1$.
Because $G$ is assumed to have an ERR:IC, Corollary~\ref{cor:erric-exist-cubic} yields that $G$ is twin-free and triangle-free; thus, $|N[u] \triangle N[v]| = 4$.
Because of the distance 4 requirement between non-detectors, at most one vertex of $N[u] \cup N[v]$ can be a non-detector, meaning $u$ and $v$ are at least 3-distinguished.

Next, suppose $d(u,v) = 2$.
Similar to the $d(u,v) = 1$ case, we know that $|N[u] \triangle N[v]| \ge 4$ (and is even).
Due to the distance 4 requirement, there can be at most one non-detector in each of $N[u]$ and $N[v]$.
If $|N[u] \triangle N[v]| \ge 6$, then we we have that $u$ and $v$ are at least 4-distinguished; otherwise, we assume $|N[u] \triangle N[v]| = 4$, implying that there is a 4-cycle $uavb$.
From lemma~\ref{lem:erric-c4-1-nondet}, we know that there is at most one non-detector in $N[u] \cup N[v]$, so $u$ and $v$ are at least 3-distinguished.

Finally, assume $d(u,v) \ge 3$.
At this distance, $N[u] \cap N[v] = \varnothing$, so $u$ and $v$ are at least 6-distinguished.
Therefore, any $\overline{S}$ satisfying these properties is an ERR:IC.

For the converse, let $S$ be an ERR:IC for $G$ and let $\overline{S} = V(G) - S$.
Lemma~\ref{lem:erric-b3} imposes the requirement that non-detectors $u,v \in \overline{S}$ have $d(u,v) \ge 4$.
Suppose that there are rivals $p,q \in V(G)$ with friends $p',q' \in V(G)$ but $\{p',q'\} \subseteq \overline{S}$; then $p$ and $q$ are not distinguished, a contradiction.
Thus, any ERR:IC must also satisfy these properties.
\end{proof}

\begin{corollary}\label{theo:erric-exist-cubic}
If $G$ is a twin-free and triangle-free cubic graph and $v \in V(G)$, then $S = V(G) - \{v\}$ is an ERR:IC for $G$.
\end{corollary}

\begin{theorem}\label{theo:erric-21-22}
If $G$ is a cubic graph with ERR:IC, then $\textrm{ERR:IC\%}(G) \le \frac{21}{22}$.
\end{theorem}
\begin{proof}
Suppose that $\overline{S} \subseteq V(G)$, satisfying the requirements of Theorem~\ref{theo:erric-sbar}, is maximal.
For any $v \in V(G)$, let $A(v) = B_3(v) \cup \{q' : v=p',q' \textrm{ are friends of rivals } p,q\}$; we see that $|A(v)| \le 22$.
The proof will proceed by showing that $\cup_{v \in \overline{S}}{A(v)} = V(G)$---that is, we cover the graph with $A(v)$ around non-detectors---and associating each detector in $A(v) \cap S$ with $v \in \overline{S}$.
Because $|A(v)| \le 22$, this gives an upper bound of $\frac{21}{22}$ for the density of $S$.

Suppose to the contrary that there is some $x \in V(G)$ such that $x \notin A(u)$ $\forall u \in \overline{S}$.
We know that $B_3(u) \subseteq A(u)$ $\forall u \in V(G)$, so it must be that $B_3(x) \subseteq S$; thus, $x$ is at least distance 4 from any non-detector.
If $x$ is not a friend with any vertex, then $\overline{S} \cup \{x\}$ still satisfies the requirements of Theorem~\ref{theo:erric-sbar}, contradicting maximality of $\overline{S}$.
Otherwise, we can assume that there are rivals $(p_1,q_1),\hdots,(p_k,q_k)$ with friends $(x,y_1),\hdots,(x,y_k)$ for some positive number $k$.
If $\{y_1,\hdots,y_k\} \subseteq S$, then $\overline{S} \cup \{x\}$ still satisfies Theorem~\ref{theo:erric-sbar}, contradicting maximality.
Otherwise, without loss of generality let $y_1 \in \overline{S}$; then $x \in A(y_1)$, a contradiction, completing the proof.
\end{proof}

\begin{figure}[ht]
    \centering
    \includegraphics[width=0.25\textwidth]{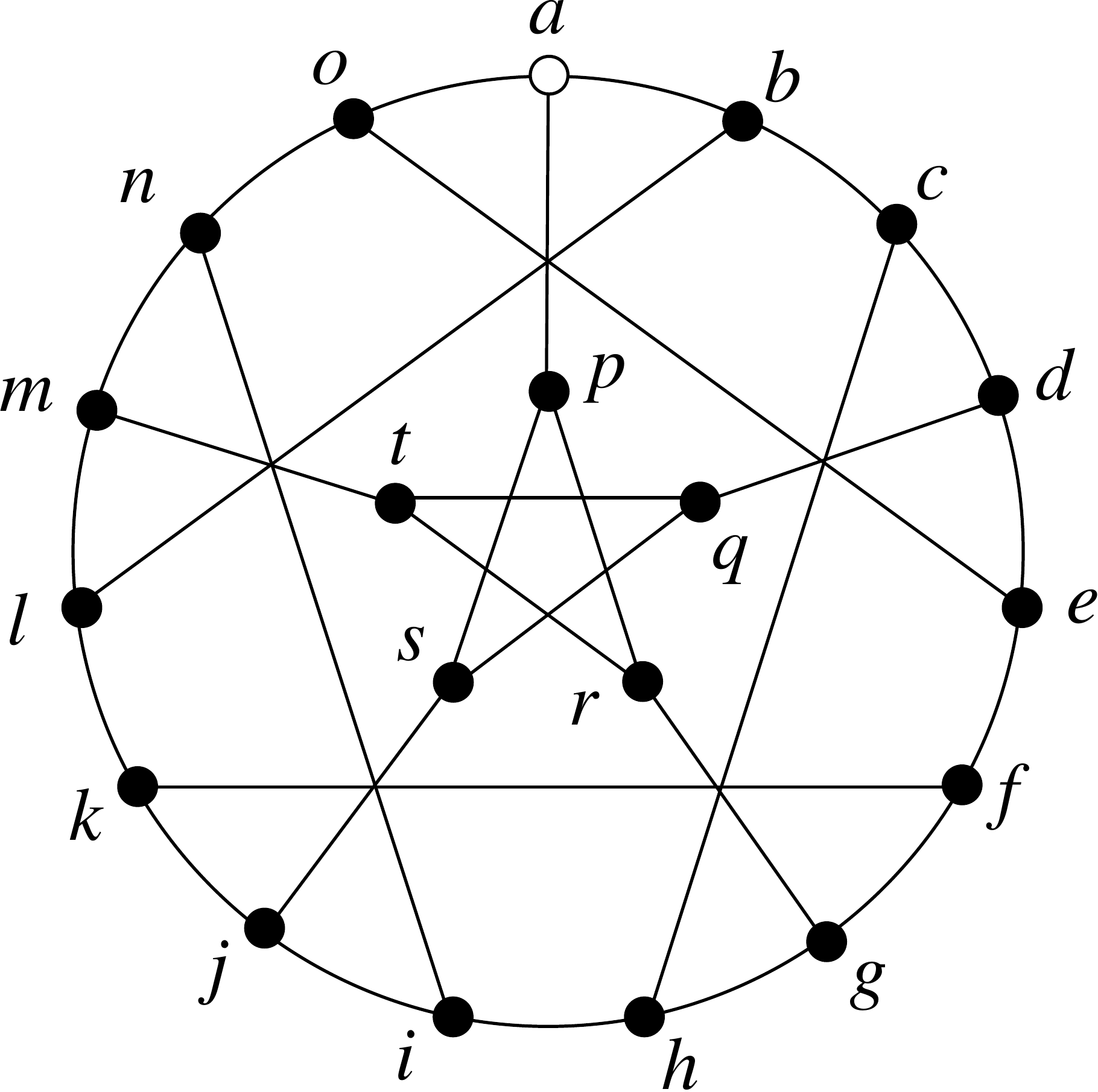}
    \caption{$\textrm{ERR:IC\%}(G_{20}) = \frac{19}{20}$}
    \label{fig:g20}
\end{figure}

Consider the graph $G_{20}$ on 20 vertices, shown in Figure~\ref{fig:g20}.
From Theorem~\ref{theo:erric-sbar}, we know that any ERR:IC on $G_{20}$ must have non-detectors separated by at least distance 4, but this is impossible on $G_{20}$ since it has diameter 3.
Thus, $\textrm{ERR:IC}(G_{20}) \ge 19$, and the solution shown in Figure~\ref{fig:g20} is indeed an ERR:IC of size 19, so $\textrm{ERR:IC}(G_{20}) = 19$.

\begin{figure}[ht]
    \centering
    \includegraphics[width=0.35\textwidth]{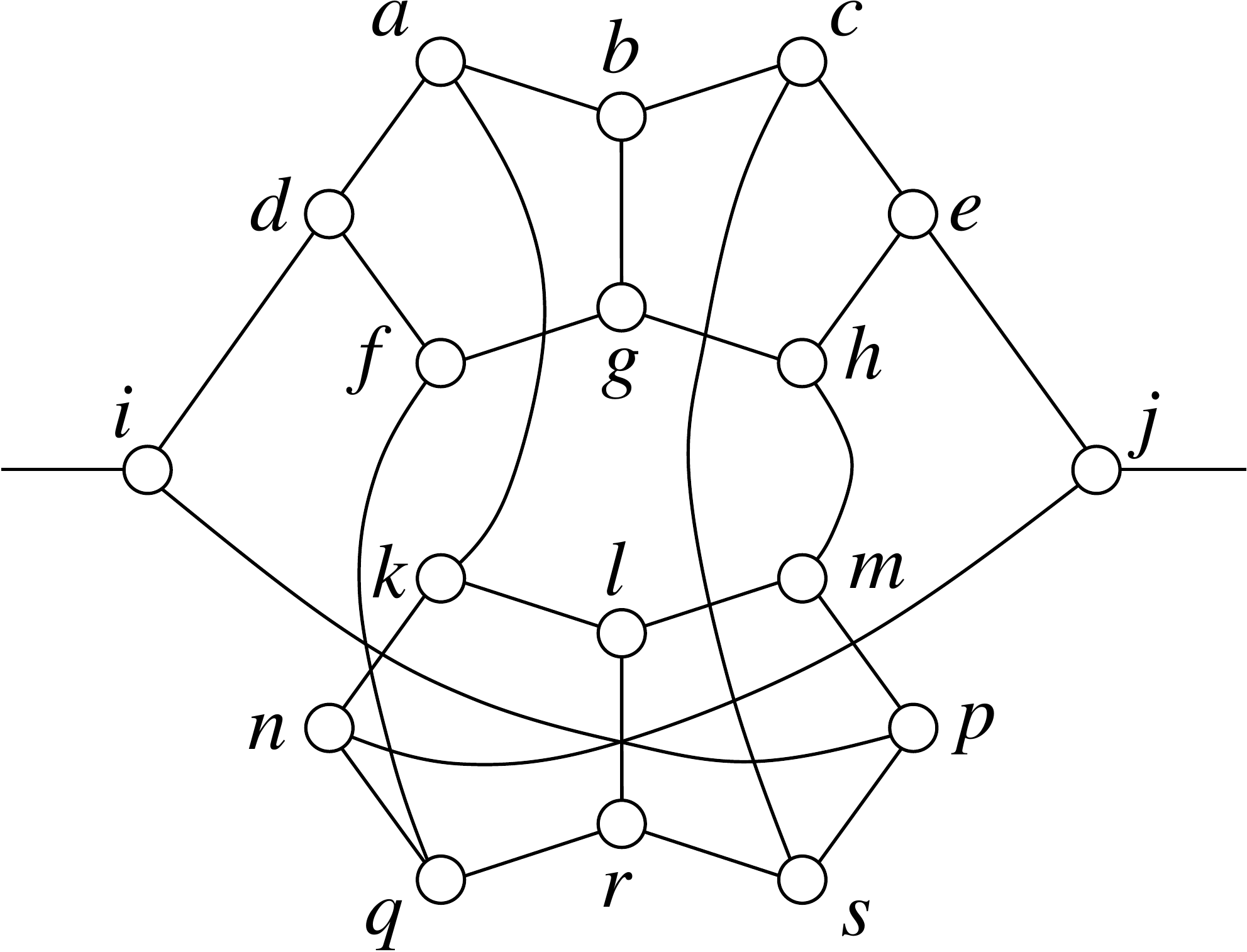}
    \caption{$G_{18}$ subgraph}
    \label{fig:g18}
\end{figure}

\begin{theorem}\label{theo:g18-fam}
The infinite family of cubic graphs given in Figure~\ref{fig:erric-11-12-fam} on $n = 18k$ vertices has $\textrm{ERR:IC}(G) = n - (k + \floor{\frac{k}{2}})$, meaning $\textrm{ERR:IC\%}(G) \ge \frac{11}{12}$.
\end{theorem}
\begin{proof}
Let $G$ be constructed by taking $k$ copies of subgraph $G_{18}$, as shown in Figure~\ref{fig:g18}, and connecting each $i$ vertex to its neighbor's $j$ vertex and vice versa.
We will show that no copy of $G_{18}$ may have more than two non-detectors, and that if a copy has two non-detectors, then neither of its neighboring copies may have two non-detectors.

Firstly, we know that if $S$ is an ERR:IC set and $x \notin S$, then $B_3(x)-\{x\} \subseteq S$.
If any vertex $x \in A = \{a,b,c,f,g,h,k,l,m,q,r,s\}$ is a non-detector, we see that $B_3(x) \subseteq V(G_{18})$, so there is at most one non-detector in that copy of $G_{18}$ and we would be done.
Otherwise, we can assume $A \subseteq S$.
The remaining vertices, $V(G_{18})-A$, are $\{d,e,i,j,n,p\}$, which we will break up into ``red" vertices $\{d,i,p\}$ and ``blue" vertices $\{e,j,n\}$.
These colors are depicted in Figure~\ref{fig:erric-11-12-fam}.

\begin{figure}[ht]
    \centering
    \includegraphics[width=0.5\textwidth]{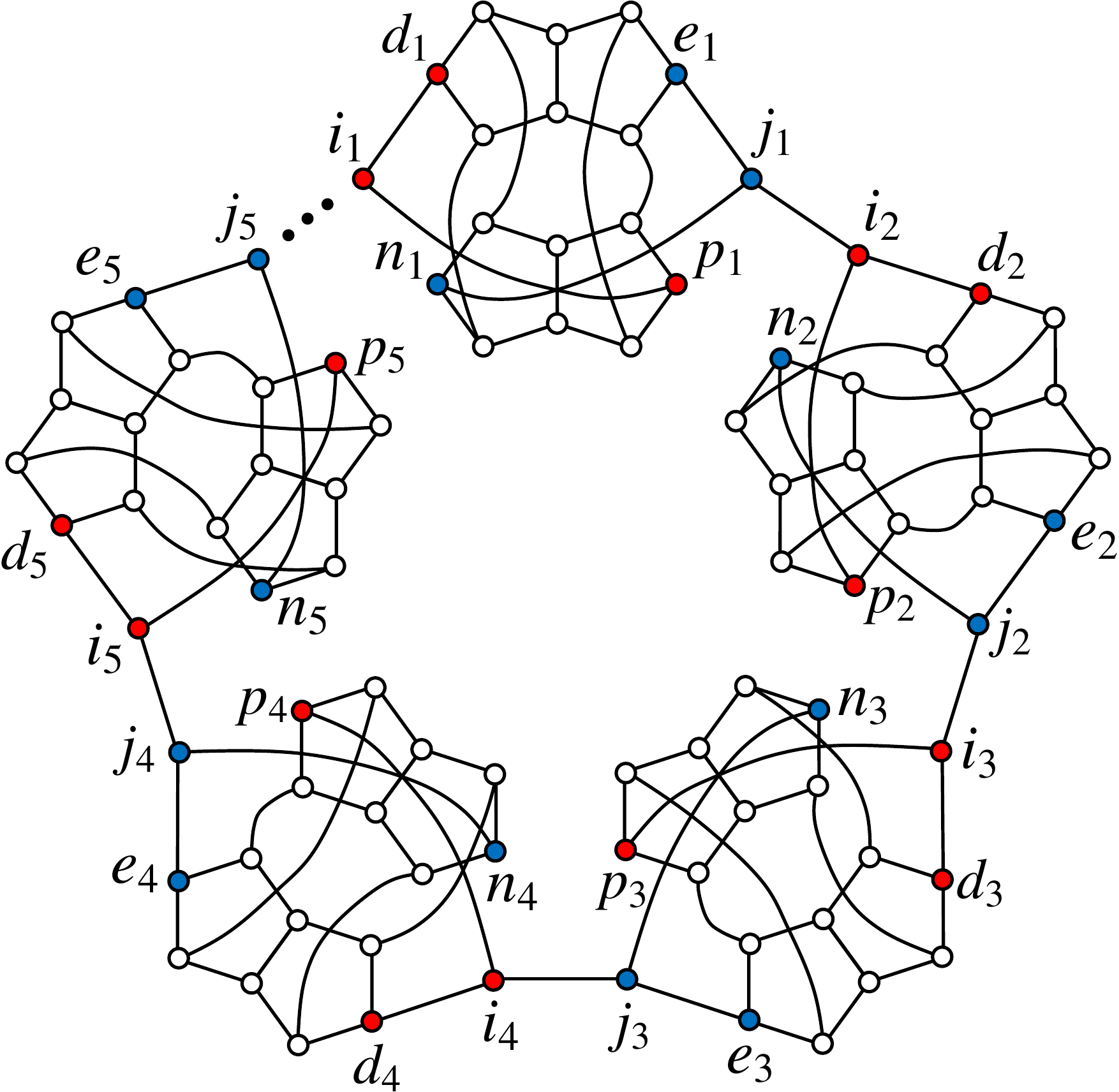}
    \caption{Infinite family of extremal cubic graph with $\textrm{ERR:IC\%}(G) \ge \frac{11}{12}$. Vertices $d$, $i$, $p$ are ``red" vertices and $e$, $j$, $n$ are ``blue" vertices.}
    \label{fig:erric-11-12-fam}
\end{figure}

We see that the red vertices are within distance 3 of one another, so at most one red vertex is a non-detector, and similarly we may have at most one blue vertex as a non-detector.
Thus, there are at most two non-detectors in $G_{18}$.
Now we assume that there are two non-detectors, one red and one blue.
By construction, the ``right" neighbor's red vertices are all within distance 3 of any of our blue vertices, and the ``left" neighbor's blue vertices are all within distance 3 of our red vertices.
Thus, if our copy has two non-detectors (a red and a blue), then the the right neighbor cannot have a red non-detector and the left neighbor cannot have a blue non-detector; thus, the neighbors both have at most one non-detector.

To get the lower bound $\textrm{ERR:IC}(G) \ge n - (k + \floor{\frac{k}{2}})$, we observe that each copy may have either 1 or 2 non-detectors, but there are no adjacent copies with 2.
Thus, as $G$ is constructed in a ring, there are at most $k + \alpha(C_k) = k + \floor{\frac{k}{2}}$ non-detectors.
To achieve this value, we can construct an $\overline{S}$ set containing the $i$ and $j$ vertices of every odd copy of $G_{18}$ (with the exception of the final copy if $k$ is odd), plus the $b$ vertices from the other copies of $G_{18}$.
This $\overline{S}$ set satisfies Theorem~\ref{theo:erric-sbar}, and so $S = V(G) - \overline{S}$ is an ERR:IC with the optimal value $\textrm{ERR:IC}(G) = n - (k + \floor{\frac{k}{2}})$.
\end{proof}

In the proof of Theorem~\ref{theo:g18-fam}, we showed that the family of cubic graphs consisting of only copies of the $G_{18}$ subgraph from Figure~\ref{fig:g18} has the property that any two adjacent copies of $G_{18}$ may not both have two non-detectors (and no copy can have three or more non-detectors).
Therefore, when $n=18$, $ij \in E(G)$ and we have only one copy of $G_{18}$ which is adjacent to itself, implying $\textrm{ERR:IC\%}(G) = \frac{17}{18}$.

From Theorem~\ref{theo:erric-21-22}, we have established that all cubic graphs which support ERR:IC can achieve a density of at most $\frac{21}{22}$.
Theorem~\ref{theo:g18-fam} then established an infinite family of cubic graphs achieving density $\frac{11}{12}$ (with odd cases having a slightly higher value), though higher values may be achievable.
Finally, we have found a single graph, shown in Figure~\ref{fig:g20}, which achieves a density of $\frac{19}{20}$; we conjecture that this is the highest value of ERR:IC of any cubic graph.

\FloatBarrier
\bibliographystyle{acm}
\bibliography{refs}

\end{document}